\documentclass[aps,prl,twocolumn,amsmath,amssymb,nofootinbib,superscriptaddress,floatfix,reprint,longbibliography]{revtex4-1}

\usepackage{amsmath} 
\usepackage{amssymb} 
\usepackage{amsthm}
\usepackage[pdftex]{color} 
\usepackage{graphicx}
\usepackage{dcolumn} 
\usepackage{bm} 
\usepackage{url}
\usepackage[colorlinks=true, urlcolor=blue, linkcolor=blue, citecolor=blue, pdftex]{hyperref}
\usepackage{longtable}
\usepackage{ulem} 
\begin{document}

\title{Chiral Phonons Coupled to Spin-Split Bands in Altermagnetic CrSb and MnTe}

\author{Armando Consiglio}
\email{consiglio@iom.cnr.it}\affiliation{CNR - Istituto Officina dei Materiali (IOM), S.S. 14, km 163.5, Trieste 34149, Italy}

\author{Maximilian Ünzelmann}
\affiliation{Experimentelle Physik VII and W\"urzburg-Dresden Cluster of Excellence ct.qmat,
Universit\"at W\"urzburg, Am Hubland, D-97074 W\"urzburg, Germany}

\author{Giancarlo Panaccione}
\affiliation{CNR - Istituto Officina dei Materiali (IOM), S.S. 14, km 163.5, Trieste 34149, Italy}

\author{Domenico Di Sante} 
\email{domenico.disante@unibo.it}
\affiliation{CNR - Istituto Officina dei Materiali (IOM), S.S. 14, km 163.5, Trieste 34149, Italy}
\affiliation{Department of Physics and Astronomy, Alma Mater Studiorum, University of Bologna, 40127 Bologna, Italy}

\date{\today}

\begin{abstract}
Altermagnets exhibit momentum-dependent spin splitting without net magnetization, providing a unique platform where magnetic order, electronic structure, and lattice dynamics intertwine. Here, using first-principles calculations, we demonstrate that CrSb and MnTe host locally chiral phonon modes carrying finite local angular momentum with a six-lobed $f$-wave texture in momentum space. The chiral lattice motion originates from the Sb/Te sublattices, whereas the altermagnetic spin splitting is generated by the magnetic Cr/Mn atoms. In pristine compounds, symmetry-related local circular polarizations cancel within each valley, leading to vanishing net phonon angular momentum. We further demonstrate that isoelectronic chemical substitution lifts this compensation and generates finite valley-resolved phonon chirality while preserving the altermagnetic character of the compounds. Most importantly, phonon calculations reveal that distortions associated with these modes produce sizable momentum-dependent modifications of the spin-split electronic bands in both materials, potentially accessible by photoemission experiments. Our results establish altermagnets as a promising platform for chiral phononics and spin-selective lattice control.
\end{abstract}

\maketitle

\setlength{\floatsep}{30pt plus 2pt minus 2pt}

{\it Introduction.} -- Chirality plays a central role across condensed matter physics, governing phenomena ranging from Weyl fermions and skyrmions to valley-selective optical excitations and topological transport phenomena \cite{ma15175812, PhysRevLett.134.226401, Bloom2024, https://doi.org/10.1002/adma.202418040, RevModPhys.90.015001, Yao2022, c2y9-3cc9, Caruso2022}. In recent years, the concept of chiral phonons has attracted growing attention \cite{Juraschek2025} due to the possibility that lattice vibrations may carry angular momentum and couple selectively to electronic, magnetic, and optical degrees of freedom. In hexagonal materials, circularly polarized phonons with quantized pseudo-angular momentum emerge at high-symmetry valleys of the Brillouin zone (BZ), producing valley-dependent selection rules and phonon Hall responses \cite{PhysRevLett.115.115502, doi:10.1126/science.aar2711, PhysRevLett.96.155901, PhysRevLett.95.155901, Cao2012, Grissonnanche2020}. Chiral phonons have been experimentally observed in transition-metal dichalcogenides through transient infrared spectroscopy \cite{doi:10.1126/science.aar2711}, in quartz via resonant inelastic X-ray scattering with circularly polarized X-rays \cite{Ueda2023}, in $\alpha$-HgS via circularly polarized Raman scattering \cite{Ishito2023}, while theoretical works have predicted analogous phenomena in kagome systems, topological materials, and magnetic lattices \cite{Slobodeniuk2023, PhysRevB.105.235204, Pan2024, Chen2023, PhysRevB.100.094303, bfll-sdrb, Tang:25, 21jc-zmzw, Zhang2025, PhysRevLett.134.196906}.

At the same time, altermagnets have recently emerged as a distinct class of collinear magnetic materials combining vanishing net magnetization with momentum-dependent spin splitting of electronic bands \cite{PhysRevX.12.040501, doi:10.1126/sciadv.aaz8809, PhysRevX.12.040002, Krempasky2024, Liu2026}. Unlike conventional antiferromagnets, altermagnets exhibit spin-polarized electronic states dictated by crystal symmetries, enabling anomalous transport, spin currents, and topological electronic phases in the absence of macroscopic magnetization. Prototypical compounds such as MnTe and CrSb have become paradigmatic platforms for the realization of altermagnetic physics due to their large momentum-dependent spin splitting and experimentally accessible electronic structure \cite{Amin2024, PhysRevLett.132.036702, PhysRevB.107.L100418, dp7v-qszq, Reimers2024, cm39-hxqk}.

Recent studies have highlighted the interplay between lattice vibrations, magnetic order, and electronic degrees of freedom, including chiral phonon responses in magnetic materials \cite{wkzr-vfgn,PhysRevLett.134.196906,PhysRevLett.134.196905,r43h-ch8r}.
In altermagnets, the occurrence of local phonon circular polarization in realistic compounds and the influence of the associated lattice distortions on momentum-dependent spin-split electronic bands remain important questions to investigate.

Here we address this problem using first-principles calculations for the prototypical altermagnets CrSb and MnTe. We demonstrate that these materials host locally chiral phonon modes at the K/K' valleys characterized by finite phonon circular polarization, which is proportional to the phonon angular momentum, and an $f$-wave momentum-space texture. Remarkably, the chiral lattice motion arises from the Sb/Te sublattice, whereas the altermagnetic electronic structure originates from the magnetic Cr/Mn atoms. In pristine compounds, inversion symmetry enforces an exact cancellation between opposite local circular motions, suppressing the total phonon angular momentum within each valley. However, symmetry lowering, induced for example by chemical substitution, removes this compensation and generates finite valley phonon angular momentum. Moreover, although in the present work we focus on chemical substitution, similar effects may naturally emerge in the pristine compound due to symmetry breaking at the surface. This is particularly relevant in CrSb and MnTe because, along the crystallographic [0001] direction, the Sb/Te atoms carrying the local circular motion occupy two distinct planes of the unit cell, so that perturbations breaking the equivalence between these layers can generate finite net phonon chirality. Most importantly, we show that these chiral phonons couple to momentum-dependent spin-split electronic bands, producing modifications of the electronic structure that can be potentially detected in Raman circular dichroism and angle-resolved photoemission spectroscopy. Our work establishes altermagnets as a natural platform where chiral lattice dynamics and spin-selective electronic structure coexist and interact.

\begin{figure}[!t]   
\includegraphics[width = 0.98\columnwidth, trim={0cm 1.3cm 0cm 0cm},clip]{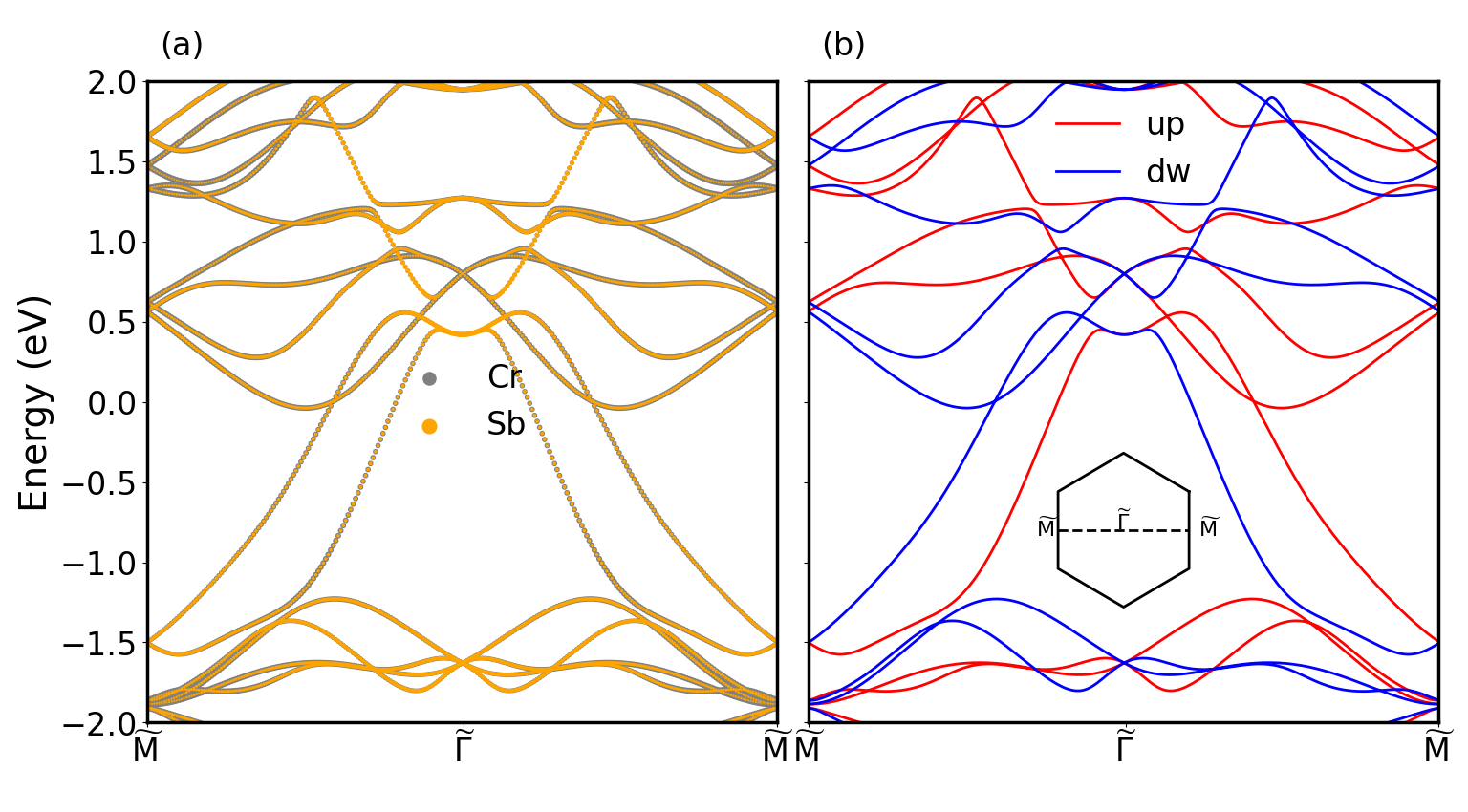}    
\caption{Electronic structure of CrSb at $k_z=0.35$.
(a) Orbital-resolved electronic bands highlighting the Cr-$d$ contribution near the Fermi level.
(b) Spin-resolved bands showing altermagnetic spin splitting. The inset indicates the momentum path in the Brillouin zone.
\label{Fig1_altermagnet}}
\end{figure}

{\it Altermagnetic electronic structure.} -- We begin by analyzing the electronic structure of CrSb in its altermagnetic ground state. A similar analysis for the MnTe compound is reported in the Supplemental Material. The magnetic order originates from the Cr sublattice, whose compensated antiparallel magnetic moments preserve vanishing net magnetization while breaking time-reversal symmetry. Consistent with previous studies \cite{Amin2024, PhysRevLett.132.036702, PhysRevB.107.L100418, dp7v-qszq, Reimers2024, cm39-hxqk}, the resulting electronic structure exhibits pronounced momentum-dependent spin splitting characteristic of $g$-wave altermagnetism. Figure \ref{Fig1_altermagnet}(a) reports the orbital-resolved band structure, highlighting the dominant Cr-$d$ contribution around the Fermi level. The corresponding spin-resolved bands shown in Fig. \ref{Fig1_altermagnet}(b) display clear spin splitting with alternating sign across the BZ, reflecting the nontrivial symmetry of the altermagnetic phase. 

The coexistence of spin-polarized electronic states and hexagonal valley structure makes CrSb an ideal candidate for investigating valley-selective lattice dynamics. As we will discuss below, the chiral phonon modes emerge predominantly from the motion of the Sb atoms, establishing a direct coupling between the nonmagnetic sublattice dynamics and the spin-split electronic states generated by the magnetic Cr atoms.

\begin{figure}[!t]   
\includegraphics[width = 0.98\columnwidth, trim={0.9cm 14.25cm 1.5cm 0.5cm},clip]{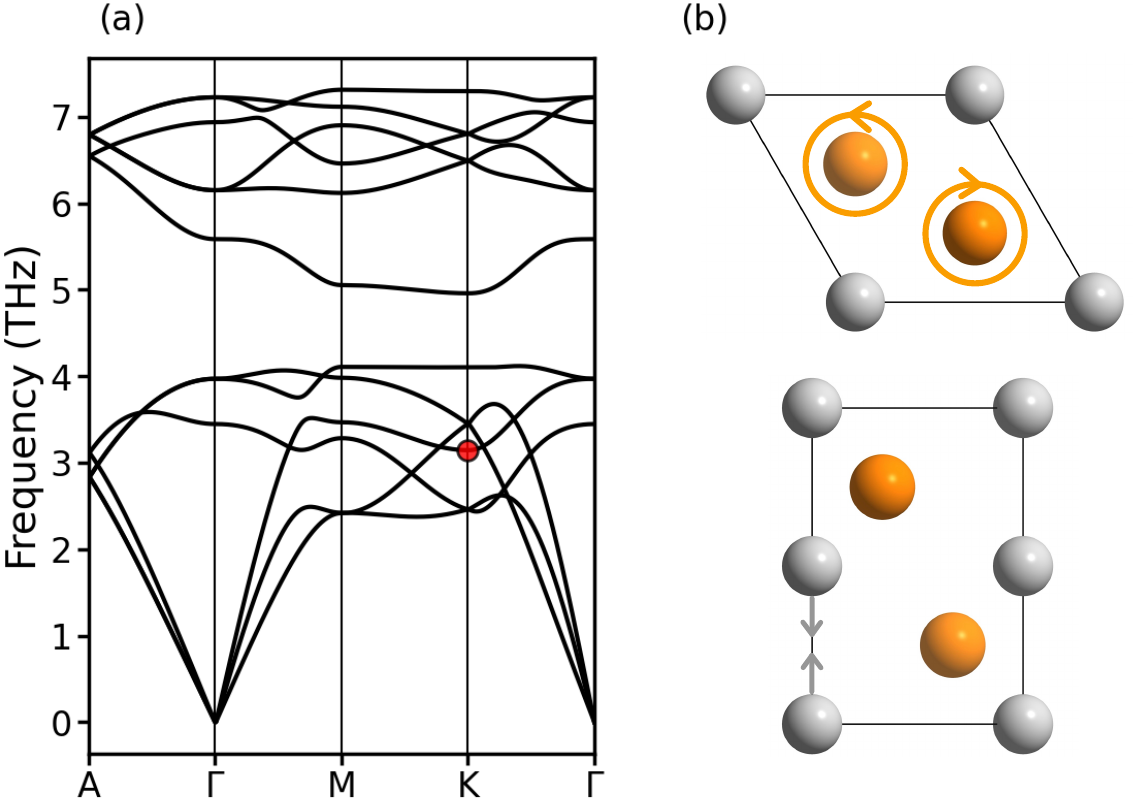}  
\caption{(a) Phonon dispersion of CrSb with a representative locally chiral K-point mode highlighted.
(b) Top and side views of its atomic motion in a $1\times1\times1$ cell: opposite circular motion of Sb atoms and linear motion of Cr atoms.
\label{Fig2_altermagnet}}
\end{figure}

{\it Local phonon circular polarization and angular momentum.} -- We investigate the lattice dynamics of CrSb using first-principles
phonon calculations (see Supplemental Material for details). The phonon dispersion in Fig.~\ref{Fig2_altermagnet}(a) reveals nondegenerate modes near the K/K' valleys exhibiting circular atomic motion in the basal plane. As illustrated in Fig.~\ref{Fig2_altermagnet}(b), this motion is carried by the Sb atoms, which rotate in opposite directions, whereas the Cr atoms remain linearly polarized.

Following Ref.~\cite{PhysRevLett.115.115502}, we quantify the local circular polarization of each phonon mode through \begin{equation}
s_{z\alpha}=\hbar\left(
|\epsilon_{R\alpha}|^2-|\epsilon_{L\alpha}|^2
\right),\qquad
s_z=\sum_\alpha s_{z\alpha},
\label{eq:operatorphononpolarization}
\end{equation}
where $\epsilon_{R\alpha}$ and $\epsilon_{L\alpha}$ are the projections of the normalized phonon eigenvector onto right- and left-circular polarization states of atom $\alpha$, respectively. The corresponding operators and polarization basis are defined in the Supplemental Material.

\begin{figure}[!t]   
\includegraphics[width = 0.98\columnwidth, trim={0cm 0.5cm 0cm 0cm},clip]{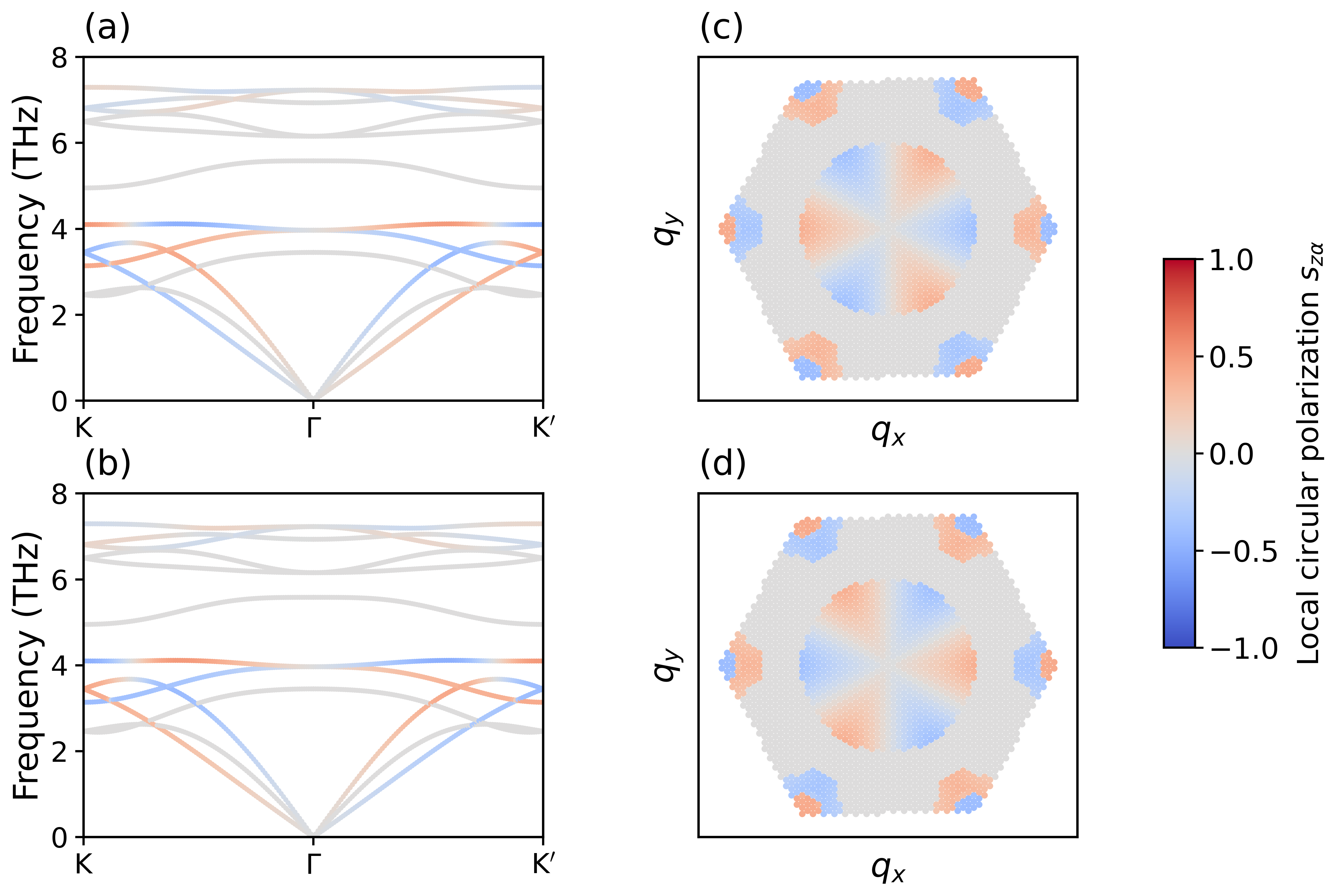} 
\caption{Local phonon circular polarization of the two Sb atoms:
(a,b) along K--$\Gamma$--K$'$ and (c,d) in the $q_z=0$ plane for the mode highlighted in Fig.~\ref{Fig2_altermagnet}(a).
The six-lobed pattern is $f$-wave-like. The color scale shows $s_{z\alpha}$ in units of $\hbar$.
\label{Fig3_altermagnet}}
\end{figure}

Figures~\ref{Fig3_altermagnet}(a,b) show finite but oppositely
signed circular polarizations on the two Sb atoms, yielding
vanishing total phonon angular momentum within each valley.
The momentum-space distributions in
Figs.~\ref{Fig3_altermagnet}(c,d) further reveal an alternating
six-lobed, $f$-wave-like texture. These results establish the
coexistence of locally chiral lattice dynamics and
altermagnetic electronic order in CrSb.

\begin{figure}[!t]   
\includegraphics[width = 0.98\columnwidth, trim={0cm 0.25cm 0cm 0.2cm},clip]{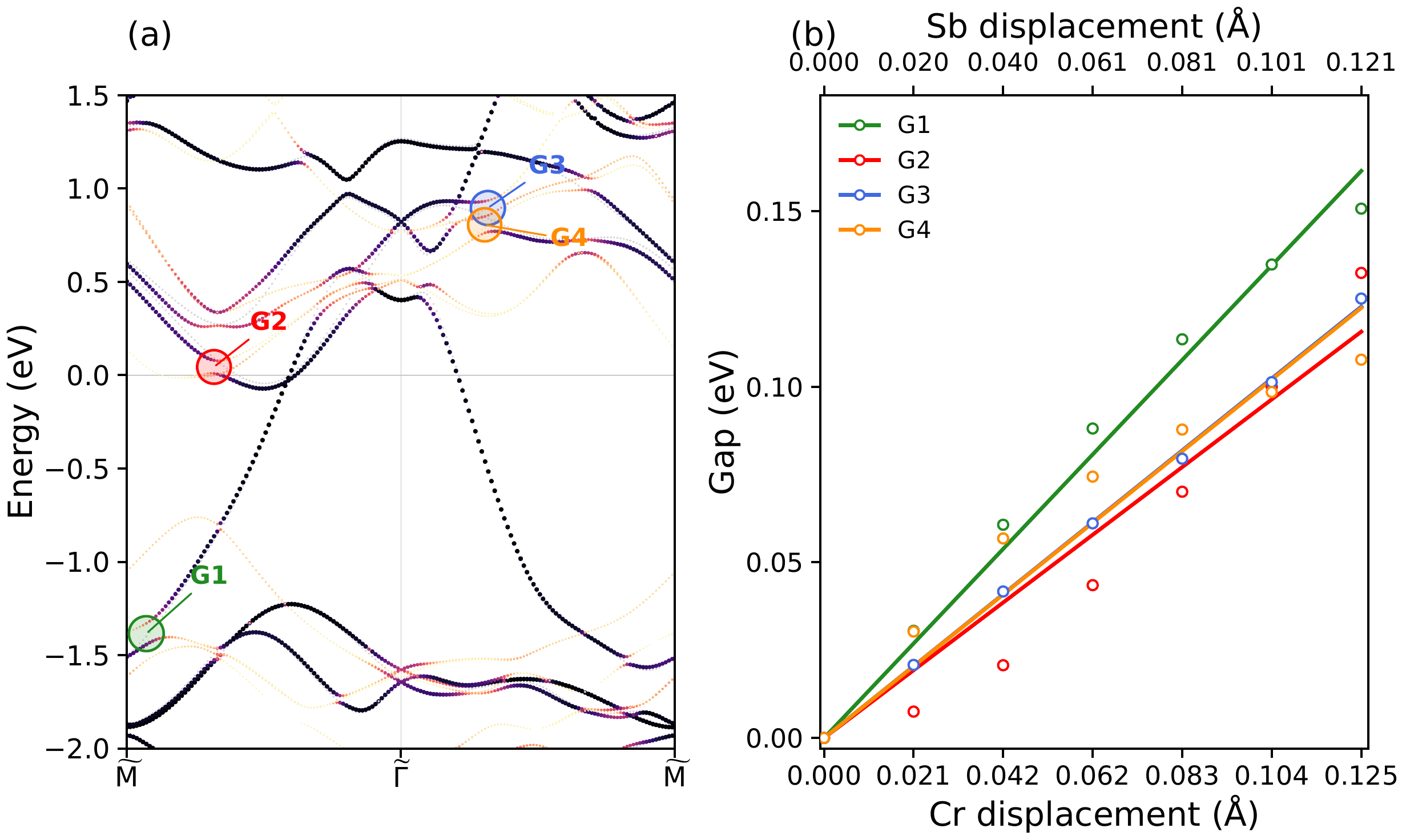}
\caption{Phonon-induced electronic reconstruction in CrSb.
(a) Unfolded bands for a frozen K-point phonon distortion at $A=0.20$~\AA{} (one spin channel, $k_z=0.35\pi/c$). Grey curves represent the undistorted bands; G1--G4 identify selected local band-repulsion regions.
(b) Corresponding local energy separations versus the maximum Cr (bottom axis) and Sb (top axis) displacements. Open circles are calculated values; solid lines are linear fits.
\label{Fig4_altermagnet}}
\end{figure}

{\it Chiral phonon-electron coupling in altermagnetic bands.} --
Having established the existence of locally chiral phonons, we now investigate how the corresponding lattice distortions affect the altermagnetic electronic structure. To this end, we construct a commensurate $3\times3\times1$ supercell and displace its atoms according to the first nondegenerate phonon mode displaying local circular polarization at the K valley. The distortion is generated from the complex phonon eigenvector, accounting for the atomic masses and the spatial phase modulation associated with the finite phonon wave vector. Here $A$ (in \AA{}) denotes the amplitude prefactor multiplying the mass-corrected phonon displacement pattern, not the maximum displacement of an individual atom (see Supplemental Material for its normalization).

Figure~\ref{Fig4_altermagnet}(a) shows the unfolded electronic band structure for a representative phonon amplitude $A=0.20$~\AA, corresponding to maximum atomic displacements of approximately $0.083$~\AA{} for Cr and $0.081$~\AA{} for Sb. The distortion produces substantial momentum-dependent changes in the spin-split electronic bands, including band repulsion and the opening of local energy gaps near crossings of the undistorted bands.

Four representative regions, labeled G1-G4 in Fig.~\ref{Fig4_altermagnet}(a), are analyzed in Fig.~\ref{Fig4_altermagnet}(b). Their local energy separations increase with the applied displacement and reach approximately $0.07$-$0.11$~eV at $A=0.20$~\AA. The approximately linear behavior is consistent with a leading-order modulation of the electronic Hamiltonian by the phonon coordinate.

To compare these distortions with a characteristic quantum fluctuation scale, we estimate the zero-point displacement contributed by the selected K-point phonon mode. For a mass-weighted phonon eigenvector normalized to unity, the mode-resolved root-mean-square displacement of atom $\alpha$ is
\begin{equation}
 u_{\alpha,0}^{(\mathbf K\nu)} =
 \sqrt{\frac{\hbar}{2N m_\alpha\omega_{\mathbf K\nu}}}\,
 \bigl|\boldsymbol\epsilon_{\alpha\nu}(\mathbf K)\bigr|,
\end{equation}
where $N=9$ is the number of primitive cells in the commensurate supercell, $m_\alpha$ is the atomic mass, and $\omega_{\mathbf K\nu}$ is the angular frequency of the selected mode. For CrSb we obtain approximately $0.0056$~\AA{} for Cr and $0.0077$~\AA{} for Sb. As a consequence of the above linear fits, we estimate local energy gaps as large as few meV. These estimates refer to the individual phonon mode at K; details of the mode normalization and its relation to the frozen-phonon amplitude $A$ are given in the Supplemental Material.

While the local phonon circular polarization is predominantly associated with Sb, the momentum-dependent spin splitting originates primarily from the magnetic Cr sublattice. The calculated band reconstruction therefore demonstrates the sensitivity of altermagnetic electronic states to a locally chiral lattice mode and suggests a possible signature in time-resolved photoemission measurements under coherent phonon excitation.

\begin{figure}[!t]   
\includegraphics[width = 0.98\columnwidth, trim={0cm 0.25cm 0cm 0.2cm},clip]{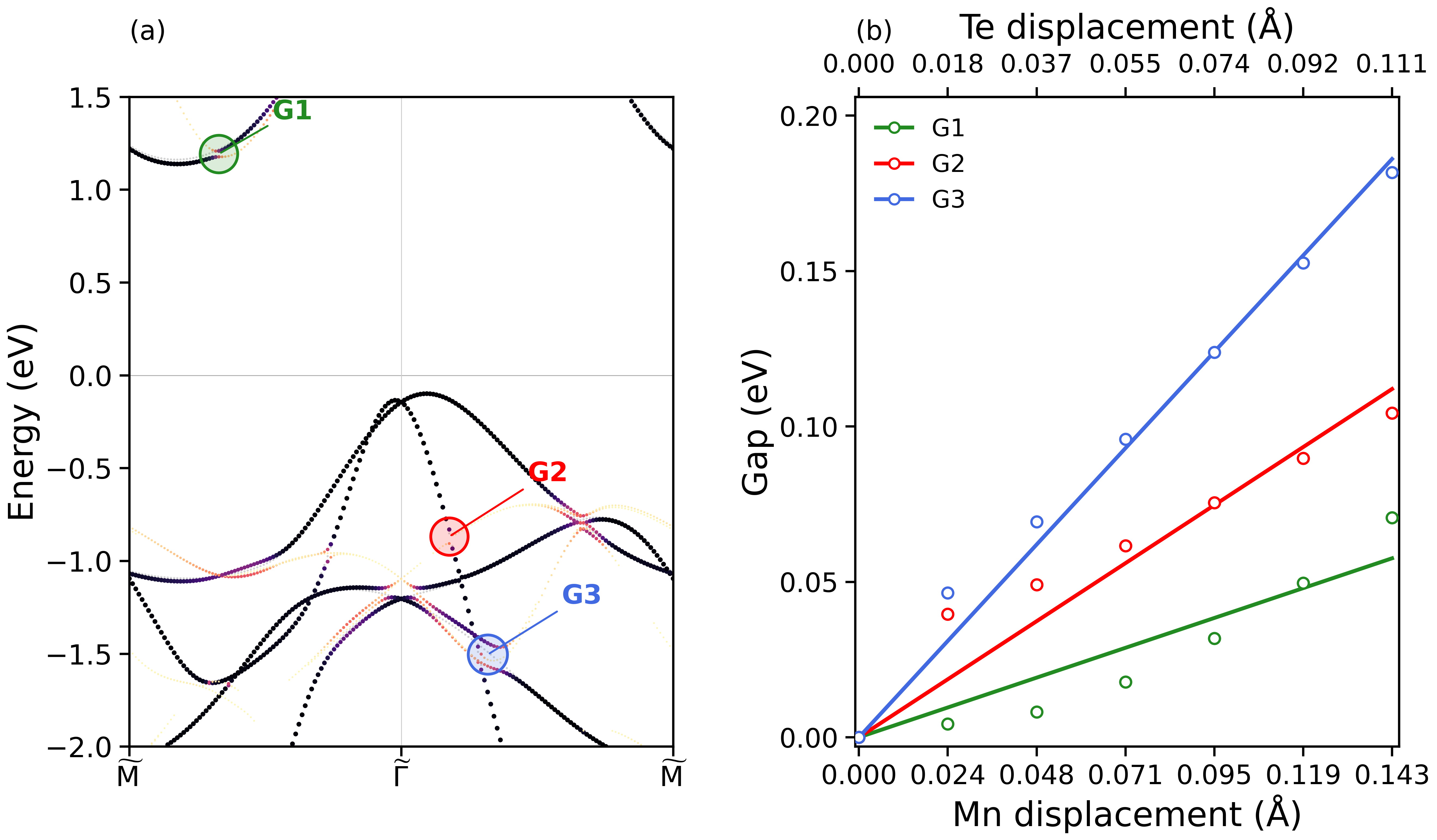}
\caption{Phonon-induced electronic reconstruction in MnTe.
(a) Unfolded bands for a frozen K-point phonon distortion at $A=0.20$~\AA{} (one spin channel). Grey curves represent the undistorted bands; G1--G3 identify selected local band-repulsion regions.
(b) Corresponding local energy separations versus the maximum Mn (bottom axis) and Te (top axis) displacements. Open circles are calculated values; solid lines are linear fits.
\label{FigMnTe_altermagnet}}
\end{figure}

A similar electronic response is found in MnTe. Figure~\ref{FigMnTe_altermagnet} shows the unfolded bands for a distortion following the locally chiral K-point phonon mode. For MnTe, we consider the phonon eigenvectors associated with the lowest-frequency mode at the K/K' valleys, using the K-point eigenvector to construct the frozen-phonon distortion. At $A=0.20$~\AA{}, the maximum Mn and Te displacements are approximately $0.095$ and $0.074$~\AA{}, respectively. Three representative local energy separations (G1--G3) reach about $0.032$, $0.076$, and $0.124$~eV, respectively, and generally increase with the displacement amplitude.\\
Thus, phonon-associated lattice distortions reconstruct the electronic bands in both altermagnets.

\begin{figure}[!t]   
\includegraphics[width = 0.98\columnwidth, trim={0cm 0.5cm 0cm 0cm},clip]{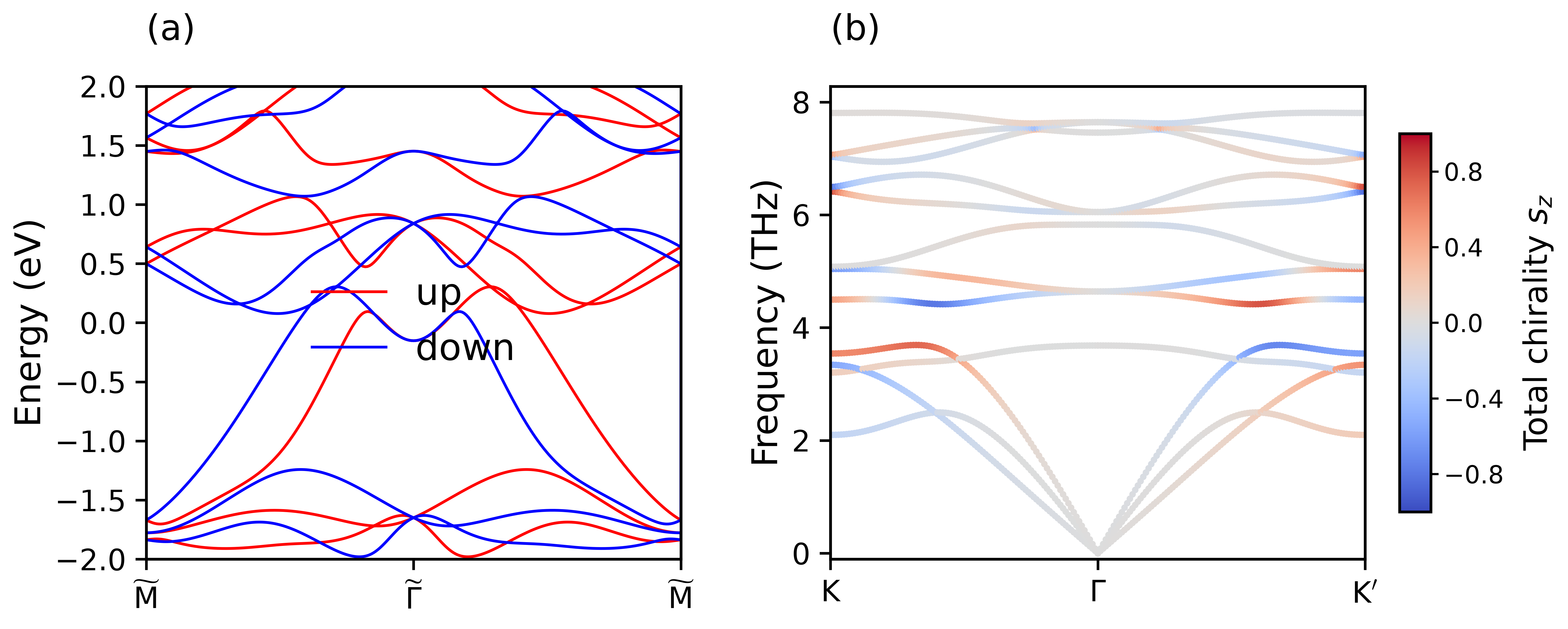}
\caption{As-substituted CrSb.
(a) Spin-resolved bands retaining the characteristic altermagnetic splitting.
(b) Phonon dispersion along K--$\Gamma$--K$'$ colored by total circular polarization $s_z$, revealing finite valley-resolved chirality.
\label{Fig5_altermagnet}}
\end{figure}

{\it Doping-induced global chirality at K/K' valleys.} -- In typical graphene-like compounds, the phonon angular momentum cancels due to symmetry-related atomic motions, resulting in vanishing net chirality within each valley \cite{PhysRevLett.115.115502}. This cancellation can be lifted by perturbations that lower the crystal symmetry. Here we investigate chemical substitution as a tunable control parameter, considering Te to Se in MnTe and Sb to As in CrSb. The considered compounds hence are Cr$_2$AsSb and Mn$_2$TeSe. These isoelectronic substitutions preserve the altermagnetic order \cite{k36v-91br} while introducing local lattice distortions and breaking inversion symmetry. 
The reduced local symmetry lifts the compensation between inequivalent atomic circular motions and generates a finite valley phonon circular polarization both at K and K'. 
To quantify the emergence of valley phonon chirality, we define the valley-resolved phonon circular polarization as
\begin{equation}
s_z^{(v)} = s_z(\mathbf{q}_v) = \sum_{\alpha}{s_{z\alpha}(\mathbf{q}_v)},
\end{equation}
where $v = \mathrm{K, K'}$ denotes the two valleys, $\mathbf{q}_v$ is the corresponding high-symmetry point and $s_{\alpha z}(\mathbf{q}_v)$ is the local circular polarization projected onto atom $\alpha$. The resulting valley contrast is then defined as
\begin{equation}
\Delta s_z = |s_z^{(\mathrm{K})} - s_z^{(\mathrm{K'})}| = 2|s_z^{(\mathrm{K/K'})}|.
\end{equation}
In the pristine compounds, symmetry enforces a complete cancellation of the phonon circular polarization within each valley, leading to $s_z^{(\mathrm{K})} = s_z^{(\mathrm{K'})} = 0$. Upon chemical substitution, this cancellation is lifted and a finite valley phonon circular polarization emerges. The two valleys are characterized by opposite chirality, $s_z^{(\mathrm{K})} = -s_z^{(\mathrm{K'})}$, resulting in the finite valley contrast $\Delta s_z$.

As shown in Fig. \ref{Fig5_altermagnet}, chemical substitution preserves the momentum-dependent spin splitting characteristic of altermagnetism, while simultaneously lifting the symmetry-protected cancellation of phonon angular momentum and leading to finite valley-resolved phonon circular polarization.
The resulting valley-selective chirality remains strongly coupled to the altermagnetic electronic structure. Because these effects originate from local symmetry breaking, we expect finite phonon chirality to survive even at moderate substitution concentrations. Interestingly, upon chemical substitution, the phonon circular polarization becomes distributed across all atomic species in the unit cell (see Supplemental Material), with finite contributions emerging on both magnetic and non-magnetic sublattices. This contrasts with the pristine case, where the chirality is predominantly localized on the Sb atoms and the Cr atoms exhibit negligible circular polarization. The same discussion applies to Se doped MnTe. Additional results for substituted structures are reported in the Supplemental Material.

{\it Conclusions.} -- We showed from first-principles that the altermagnets CrSb and MnTe host local chiral phonon modes carrying finite phonon angular momentum and $f$-wave momentum-space textures. The chiral lattice dynamics originate from the Sb/Te sublattices, whereas the momentum-dependent spin splitting characteristic of altermagnetism is generated by the magnetic Cr/Mn atoms. This demonstrate that, locally, chiral phonon distortions can directly affect the momentum-dependent electronic structure of altermagnetic CrSb and MnTe. Frozen-phonon displacements associated with locally chiral modes produce measurable modifications of the spin-split electronic structure, demonstrating a strong interplay between lattice dynamics and altermagnetic electronic states.
Although pristine compounds exhibit vanishing net valley chirality due to symmetry-enforced cancellation, isoelectronic chemical substitution lifts this compensation and induces finite phonon angular momentum at K/K' valleys, while preserving the underlying altermagnetic nature of the compounds. As a result, a finite net valley phonon chirality emerges, providing a tunable route to convert local chiral lattice dynamics into an observable macroscopic response. Furthermore, finite valley phonon chirality may emerge in undoped compounds through symmetry breaking induced by surfaces. Since the Sb/Te atoms carrying the local circular polarization are located on distinct planes of the unit cell, perturbations that break the equivalence between these layers may naturally convert the compensated local chiral motion into a finite observable phonon angular momentum.
Our results identify altermagnets as a promising platform for chiral phononics, where spin, valley, and lattice degrees of freedom can be manipulated simultaneously in the absence of macroscopic magnetization. More broadly, the outcome suggests that chiral lattice dynamics may provide a route toward controlling nonequilibrium spin textures and valley-selective electronic responses in magnetic quantum materials.

{\it Acknowledgments.} 
 A.C. acknowledges support from PNRR MUR Project No. PE0000023-NQSTI. A.C. and D.D.S. acknowledge the Gauss Centre for Supercomputing e.V. for funding this project by providing computing time on the GCS Supercomputer SuperMUC-NG at Leibniz Supercomputing Centre. D.D.S acknowledges MUR funding within the FIS2 (n. 1236, 01-08-2023) Project no. FIS-2023-00144 (CUP J53C25001880001). A.C. and D.D.S. acknowledge useful discussions with Alberto Guandalini and Paolo Barone. 

\bibliography{bibliography}

\end{document}


\title{Supplemental Material for\\
``Chiral Phonons Coupled to Spin-Split Bands in Altermagnetic CrSb and MnTe''}

\author{Armando Consiglio}
\email{consiglio@iom.cnr.it}
\affiliation{CNR - Istituto Officina dei Materiali (IOM), S.S. 14, km 163.5, Trieste 34149, Italy}

\author{Maximilian \"Unzelmann}
\affiliation{Experimentelle Physik VII and W\"urzburg-Dresden Cluster of Excellence ct.qmat,
Universit\"at W\"urzburg, Am Hubland, D-97074 W\"urzburg, Germany}

\author{Giancarlo Panaccione}
\affiliation{CNR - Istituto Officina dei Materiali (IOM), S.S. 14, km 163.5, Trieste 34149, Italy}

\author{Domenico Di Sante}
\email{domenico.disante@unibo.it}
\affiliation{CNR - Istituto Officina dei Materiali (IOM), S.S. 14, km 163.5, Trieste 34149, Italy}
\affiliation{Department of Physics and Astronomy, Alma Mater Studiorum, University of Bologna, 40127 Bologna, Italy}

\maketitle

\renewcommand{\thefigure}{S\arabic{figure}}
\renewcommand{\theequation}{S\arabic{equation}}
\renewcommand{\thetable}{S\arabic{table}}
\setlength{\textfloatsep}{12pt plus 2pt minus 2pt}
\setlength{\floatsep}{12pt plus 2pt minus 2pt}

\section{Computational Methods}

\subsection{Density-functional theory and structural relaxation}

First-principles calculations were performed within density functional theory (DFT) using the Vienna \textit{Ab initio} Simulation Package (VASP)~\cite{KresseFurthmuller1996PRB,KresseFurthmuller1996CMS}. The exchange-correlation functional was treated within the Perdew--Burke--Ernzerhof (PBE) generalized-gradient approximation~\cite{PerdewBurkeErnzerhof1996}, using projector-augmented-wave (PAW) potentials~\cite{Blochl1994PAW,KresseJoubert1999PAW}.

Electronic correlations in the transition-metal $d$ orbitals were included within the rotationally invariant DFT+$U$ approach of Dudarev \textit{et al.}~\cite{Dudarev1998DFTU}. An effective Hubbard parameter of $U_{\mathrm{eff}}=0.8$ eV was applied to the Cr-$d$ orbitals in CrSb, while $U_{\mathrm{eff}}=4.8$ eV was employed for the Mn-$d$ orbitals in MnTe. No Hubbard correction was applied to the pnictogen or chalcogen states.

The primitive cells of CrSb and MnTe were fully relaxed using a plane-wave kinetic-energy cutoff of 600 eV and a $12\times12\times12$ $\mathbf{k}$-point mesh. Gaussian smearing with a width of 0.02 eV was used. Both lattice parameters and internal atomic coordinates were optimized. The electronic self-consistency threshold was set to $10^{-8}$ eV. The structural optimizations were performed using a conjugate-gradient ionic relaxation with both ionic coordinates and cell degrees of freedom allowed to relax.

The compensated collinear magnetic configuration was initialized by assigning opposite magnetic moments of $+5\,\mu_{\mathrm B}$ and $-5\,\mu_{\mathrm B}$ to the two transition-metal atoms in the primitive cell, with zero initial magnetic moments on the Sb/Te atoms.


\subsection{Phonon calculations}

Phonon calculations were performed within VASP using the finite-displacement approach~\cite{Parlinski1997Direct,TogoTanaka2015Phonopy}. A $3\times3\times3$ supercell of the primitive cell was employed together with a $3\times3\times3$ electronic $\mathbf{k}$-point mesh. The displacement amplitude was set to 0.015~\AA. The electronic self-consistency threshold was $10^{-8}$ eV and the calculations were performed using the \texttt{Accurate} precision setting.

A cutoff of 800 eV was used for both compounds. Gaussian smearing with a width of 0.05 eV was employed for the supercell calculations. The same DFT+$U$ parameters used for the primitive cells were retained, namely $U_{\mathrm{eff}}=0.8$ eV for Cr-$d$ states and $U_{\mathrm{eff}}=4.8$ eV for Mn-$d$ states. Nonspherical contributions to the gradient corrections inside the PAW spheres were included.

The compensated magnetic configuration was extended to the $3\times3\times3$ supercells by assigning opposite initial magnetic moments of $\pm5\,\mu_{\mathrm B}$ to the two magnetic sublattices and zero initial moments to the pnictogen/chalcogen sites.

The real-space force constants obtained from the VASP finite-displacement calculations were subsequently post-processed with Phonopy~\cite{TogoTanaka2015Phonopy,Togo2023Phonopy}. Phonopy was used to construct and Fourier-interpolate the dynamical matrix and to obtain the phonon frequencies and complex, mass-weighted eigenvectors along the selected reciprocal-space paths and momentum-space meshes. The complex eigenvectors were used directly for the analysis of the local phonon circular polarization and angular momentum described below.

\subsection{Electronic band unfolding and local band separations}

For the commensurate frozen-phonon configurations described below, static electronic calculations were performed and the resulting supercell bands were unfolded to the primitive Brillouin zone using VASPKIT~\cite{Wang2021VASPKIT}. The plotted spectral weight measures the primitive-cell Bloch character of each supercell eigenstate. The reported G1-G4 (CrSb) and
G1-G3 (MnTe) are representative local energy separations in regions of phonon-induced band repulsion. The drawn straight lines are fits constrained through the origin and serve to illustrate the leading response to the structural amplitude. These calculations establish band reconstruction associated with the selected lattice mode.

\subsection{Phonon circular polarization and angular momentum}
We analyze the phonon modes using the complex, mass-weighted eigenvectors $|\boldsymbol\epsilon_{\nu\mathbf q}\rangle$ of the dynamical matrix, normalized as $\sum_{\alpha i}|\epsilon_{\alpha i,\nu}(\mathbf q)|^2=1$. The atom order follows the primitive cells $(\mathrm{Cr/Mn}_1,\mathrm{Cr/Mn}_2,
\mathrm{Sb/Te}_1,\mathrm{Sb/Te}_2)$. For each atom $\alpha$, it is possible to introduce the
normalized in-plane circular basis states:
\begin{equation}
 |R_\alpha\rangle=\tfrac{1}{\sqrt{2}}(\ldots,1,i,0,\ldots)^T,\qquad
 |L_\alpha\rangle=\tfrac{1}{\sqrt{2}}(\ldots,1,-i,0,\ldots)^T.
 \label{eq:SM_RL}
\end{equation}
Only the components of atom $\alpha$ are nonzero in \ref{eq:SM_RL}, in the local Cartesian
$(x,y,z)$ order. The atom-resolved polarization operator is:
\begin{equation}
 \hat S_{z\alpha}=|R_\alpha\rangle\langle R_\alpha|
                    -|L_\alpha\rangle\langle L_\alpha|.
 \label{eq:SM_Sz}
\end{equation}
With $\epsilon_{R\alpha}=\langle R_\alpha|\boldsymbol\epsilon_{\nu\mathbf q}\rangle$
and $\epsilon_{L\alpha}=\langle L_\alpha|\boldsymbol\epsilon_{\nu\mathbf q}\rangle$,
we obtain:
\begin{equation}
 \frac{s_{z\alpha}(\nu,\mathbf q)}{\hbar}
 =|\epsilon_{R\alpha}|^2-|\epsilon_{L\alpha}|^2
 =2\,\mathrm{Im}(\epsilon_{\alpha x}^{*}\epsilon_{\alpha y}),
 \label{eq:SM_localchirality}
\end{equation}
\begin{equation}
 s_z(\nu,\mathbf q)=\sum_\alpha s_{z\alpha}(\nu,\mathbf q).
 \label{eq:SM_totalchirality}
\end{equation}
The sign follows the circular-basis convention in Eq.~\eqref{eq:SM_RL}. The color scales in the main text show $s_{z\alpha}/\hbar$ or $s_z/\hbar$, respectively; no normalization by the largest value in the data is applied. In particular, we do not divide the local polarization by $|\epsilon_{\alpha x}|^2+|\epsilon_{\alpha y}|^2$, since that ratio measures local circularity rather than the atom's weighted contribution to the phonon polarization. Consequently, a strongly circular atomic trajectory can still contribute little if its displacement weight in the normalized
phonon eigenvector is small.

A finite $s_{z\alpha}$ does not imply a finite total $s_z$. For the selected modes of pristine CrSb and MnTe, the computed Sb/Te contributions carry opposite signs and cancel when summed over the primitive cell. The comparison is made mode by mode at fixed $\mathbf q$.

\subsection{Momentum-resolved phonon chirality}

For the plots in the main text, the polarization $s_{z\alpha}(\nu,\mathbf q)$ was evaluated along the $\mathrm K$-$\Gamma$-$\mathrm K'$ path and throughout the $q_z=0$ plane. For a mesh specified in reciprocal fractional coordinates, $\mathbf q=q_1\mathbf b_1+q_2\mathbf b_2+q_3\mathbf b_3$ was transformed to Cartesian coordinates before plotting the two-dimensional maps. The atom projections are obtained by evaluating Eq.~\eqref{eq:SM_localchirality} for each normalized mode; the same dimensionless polarization is represented in the band plots and the two-dimensional maps.

The CrSb maps in the main text use the selected phonon branch identified in the phonon dispersion. For MnTe, the K-point frozen-phonon calculation uses the lowest-frequency, nondegenerate mode of frequency $f_{\mathrm K}\approx  1.92$~THz. A branch index in the plots is assigned after ordering the eigenfrequencies at a given wave vector; tracking a particular mode across accidental crossings may require matching eigenvectors, not merely sorting the frequencies. The opposite local contributions of Sb/Te produce the alternating six-lobed textures reported in the main text.

\subsection{50\% chemically substituted compounds}

To investigate the effect of symmetry breaking on the phonon circular polarization, we additionally considered ordered structures with 50\% isoelectronic substitution of the pnictogen/chalcogen sublattice. Specifically, one of the two Sb sites in CrSb was replaced by As, yielding Cr$_2$SbAs, while one of the two Te sites in MnTe was replaced by Se, yielding Mn$_2$TeSe. These structures therefore correspond to 50\% substitution on the Sb/Te sublattice.

The substituted compounds were treated using the same first-principles methodology employed for the corresponding pristine systems. The same magnetic configurations and Hubbard parameters were retained. Their phonon properties were calculated using $3\times3\times3$ supercells and $3\times3\times3$ electronic $\mathbf{k}$-point meshes, following the same finite-displacement procedure used for the pristine compounds.

At 50\% substitution, the two pnictogen/chalcogen sites that are symmetry equivalent in pristine CrSb and MnTe become chemically inequivalent. This removes the symmetry responsible for the exact compensation of their local circular polarizations and allows a finite total phonon circular polarization to emerge at the K and K$'$ valleys.

The valley-resolved circular polarization was evaluated as
\begin{equation}
s_z^{(v)}
=
s_z(\mathbf{q}_v)
=
\sum_\alpha s_{z\alpha}(\mathbf{q}_v),
\qquad v=\mathrm{K,K'},
\end{equation}
and the corresponding valley contrast was defined as
\begin{equation}
\Delta s_z
=
\left|
s_z^{(\mathrm K)}
-
s_z^{(\mathrm K')}
\right|.
\end{equation}

\subsection{Frozen-phonon displacement pattern and zero-point scale}

The electronic structures of the pristine and ordered substituted compounds were obtained with VASP using the DFT+$U$ settings described above. To study phonon-induced band reconstruction, we froze a commensurate real-space configuration associated with a single selected complex K-point phonon eigenvector. The supercells contain $3\times3\times1$ primitive cells ($N=9$), with $\mathbf K=(1/3,1/3,0)$ in reciprocal fractional
coordinates.

\subsubsection{Displacement amplitude and phase convention}

We start from normalized mass-weighted eigenvectors $\boldsymbol\epsilon_{\alpha\nu}(\mathbf K)$ and introduce the mode-dependent effective mass
\begin{equation}
 M_{\mathrm{eff}}^{-1}=\sum_\alpha
   \frac{|\boldsymbol\epsilon_{\alpha\nu}(\mathbf K)|^2}{m_\alpha},
 \qquad \sum_\alpha|\boldsymbol\epsilon_{\alpha\nu}(\mathbf K)|^2=1.
 \label{eq:SM_Meff}
\end{equation}
The displacement of atom $\alpha$ in cell $\ell$ is parameterized as:
\begin{equation}
 \mathbf u_{\ell\alpha}(A,\theta)=
 A\,\mathrm{Re}\!\left[
 \sqrt{\frac{M_{\mathrm{eff}}}{m_\alpha}}\,
 \boldsymbol\epsilon_{\alpha\nu}(\mathbf K)
 e^{i[\mathbf K\cdot(\mathbf R_\ell+\boldsymbol\tau_\alpha)-\theta]}
 \right].
 \label{eq:SM_frozen}
\end{equation}
Here $\mathbf R_\ell$ is a primitive-cell lattice translation, $\boldsymbol\tau_\alpha$ is the atomic position within that cell, and $\theta$ selects the phase (snapshot) of the displacement. The dot product is evaluated with Cartesian reciprocal and real-space vectors, or equivalently as $2\pi$ times the fractional-coordinate product. The factor $\exp(i\mathbf K\cdot\boldsymbol\tau_\alpha)$ is included explicitly in our POSCAR generators. Equation~\eqref{eq:SM_frozen} describes a real static configuration constructed from the complex K-point eigenvector.

The scalar $A$ has units of \AA{} and multiplies the complete mass-corrected mode pattern. It is not the maximum displacement of an atom: that value depends on the species, eigenvector and phase $\theta$. For the chosen snapshot $\theta=0$ and $A=0.20$~\AA, the calculated largest atomic displacements are $0.0831$~\AA{} (Cr), $0.0809$~\AA{} (Sb), $0.0951$~\AA{} (Mn), and $0.0740$~\AA{} (Te). The effective masses for the two selected modes are approximately $97.67$~u (CrSb) and $94.72$~u (MnTe). These values and conventions are used consistently in the static supercells and in the plots of the band separations versus atomic displacement.

\subsubsection{Single-mode zero-point scale}

For comparison with the static frozen-phonon distortions, we estimate the
zero-point scale associated with the real normal coordinate generated by the
selected commensurate $K$-point mode. For the displacement convention of
Eq.~\eqref{eq:SM_frozen}, we define the corresponding mass-weighted normal
coordinate $Q$ through
\begin{equation}
Q^2
=
\sum_{\ell\alpha}
m_\alpha
\left|
\mathbf u_{\ell\alpha}
\right|^2 .
\label{eq:SM_Qdef}
\end{equation}
For the $3\times3\times1$ commensurate supercell, the spatial average of the
squared real part of the complex $K$-point eigenvector gives
\begin{equation}
Q^2
=
\frac{N}{2}
M_{\mathrm{eff}}A^2,
\qquad N=9.
\label{eq:SM_QA}
\end{equation}
Thus the frozen-phonon amplitude $A$ is directly related to the real normal
coordinate $Q$ associated with the selected $K/K'$ pair.

Within the harmonic approximation, this normal coordinate is described by
\begin{equation}
\hat H_{\mathrm{ph}}
=
\frac{1}{2}\hat P^2
+
\frac{1}{2}
\omega_{\mathbf K\nu}^{2}\hat Q^2 ,
\end{equation}
where $\omega_{\mathbf K\nu}=2\pi f_{\mathbf K\nu}$. In the vibrational ground
state,
\begin{equation}
\left\langle \hat Q^2 \right\rangle_0
=
\frac{\hbar}{2\omega_{\mathbf K\nu}}.
\label{eq:SM_Qzpm}
\end{equation}
Combining Eqs.~\eqref{eq:SM_QA} and \eqref{eq:SM_Qzpm}, the root-mean-square
zero-point value of the frozen-phonon amplitude is
\begin{equation}
A_0
=
\sqrt{
\frac{\hbar}
{N M_{\mathrm{eff}}\omega_{\mathbf K\nu}}
}.
\label{eq:SM_Azpm}
\end{equation}

Using Eq.~\eqref{eq:SM_frozen}, the corresponding root-mean-square
zero-point displacement of atom $\alpha$ is
\begin{equation}
u_{\alpha,0}^{(\mathbf K\nu)}
=
A_0
\sqrt{\frac{M_{\mathrm{eff}}}{m_\alpha}}
\frac{
\left|
\boldsymbol\epsilon_{\alpha\nu}(\mathbf K)
\right|
}{\sqrt{2}},
\end{equation}
which gives
\begin{equation}
\boxed{
u_{\alpha,0}^{(\mathbf K\nu)}
=
\sqrt{
\frac{\hbar}
{2N m_\alpha\omega_{\mathbf K\nu}}
}\,
\left|
\boldsymbol\epsilon_{\alpha\nu}(\mathbf K)
\right|
}.
\label{eq:SM_zpm}
\end{equation}
The factor $1/\sqrt{N}$ reflects the normalization of the extended phonon mode
over the $N$ primitive cells, while the additional factor $1/\sqrt{2}$ arises
from the spatial average of the real part of the complex $K$-point eigenvector
over the commensurate supercell.

For CrSb, this yields approximately $0.0056$~\AA{} for Cr and
$0.0077$~\AA{} for Sb. These values represent the zero-point scale associated
with the selected real normal coordinate constructed from the $K/K'$ pair;
they are not the total zero-point mean-square displacement of the atom, which
would contain contributions from all phonon wave vectors and branches.
Furthermore, they should not be identified with the externally imposed
coherent frozen-phonon amplitude $A$ used in the static calculations.


\subsubsection{Harmonicity check from frozen-phonon total energies}

As an independent consistency check on the frozen-phonon amplitudes, we calculated the DFT total energy of the same commensurate real-space distortions used for the electronic-structure analysis.

For the displacement convention introduced above,
\begin{equation}
\mathbf u_{\ell\alpha}(A,\theta)
=
A\,\mathrm{Re}\!\left[
\sqrt{\frac{M_{\mathrm{eff}}}{m_\alpha}}\,
\boldsymbol\epsilon_{\alpha\nu}(\mathbf K)
e^{i[\mathbf K\cdot(\mathbf R_\ell+\boldsymbol\tau_\alpha)-\theta]}
\right],
\end{equation}
with the phonon eigenvector normalized according to
\begin{equation}
\sum_{\alpha}
\left\|
\boldsymbol\epsilon_{\alpha\nu}(\mathbf K)
\right\|^2
=
1.
\end{equation}

For a commensurate K-point distortion in a $3\times3\times1$
supercell, the phase factors sampled by the real part of the complex
eigenvector satisfy
\begin{equation}
\frac{1}{N}
\sum_{\ell}
\cos^2\phi_{\ell\alpha}
=
\frac{1}{2},
\end{equation}
where $N=9$ is the number of primitive cells in the in-plane
supercell and
\begin{equation}
\phi_{\ell\alpha}
=
\mathbf K\cdot
(\mathbf R_\ell+\boldsymbol\tau_\alpha)
-\theta .
\end{equation}
Consequently,
\begin{equation}
\sum_{\ell\alpha}
m_\alpha
\left|
\mathbf u_{\ell\alpha}
\right|^2
=
\frac{N}{2}
M_{\mathrm{eff}}A^2.
\end{equation}

Within the harmonic approximation, the potential-energy cost of the
frozen distortion is therefore:
\begin{equation}
\Delta E_{\mathrm{harm}}(A)
=
\frac{1}{2}
\omega_{\mathbf K\nu}^{2}
\sum_{\ell\alpha}
m_\alpha
\left|
\mathbf u_{\ell\alpha}
\right|^2
=
\frac{N}{4}
M_{\mathrm{eff}}
\omega_{\mathbf K\nu}^{2}
A^2.
\label{eq:SM_harmonic_energy}
\end{equation}
Thus, the factor $1/4$ is specific to the present definition of the
distortion amplitude $A$: one factor $1/2$ is the usual harmonic
potential-energy prefactor, while the second arises from the spatial
average of the squared real part of the complex K-point eigenvector
over the commensurate supercell.

For CrSb, the selected mode has $M_{\mathrm{eff}}=97.6747$~u and $f_{\mathbf K}=3.139713$~THz, which gives $c_{\mathrm{harm}}=8.8643$~eV/\AA$^2$. The calculated total-energy differences for $A=0,0.05,\ldots,0.30$~\AA{} are $0$, $21.98$, $88.44$, $199.85$, $355.53$, $555.47$, and $801.39$~meV, respectively. A quadratic fit constrained to pass through the undistorted structure gives:
\begin{equation}
 \Delta E_{\mathrm{DFT}}^{\mathrm{CrSb}}(A) = 8.8966\,A^{2}\ {\rm eV},
 \label{eq:SM_harmonic_CrSb}
\end{equation}
only $0.36\%$ above the harmonic prediction. Equivalently, the curvature corresponds to $f_{\mathrm{DFT}}\simeq3.1454$~THz, in excellent agreement with the phonon frequency $3.1397$~THz. Thus the selected CrSb distortion remains essentially harmonic over the full amplitude range considered.

For MnTe, the selected mode has $M_{\mathrm{eff}}=94.7208$~u and $f_{\mathbf K}=1.922131$~THz, giving $c_{\mathrm{harm}}=3.22175$~eV/\AA$^2$. The DFT total-energy differences for $A=0,0.05,\ldots,0.30$~\AA{} are $0$, $7.73$, $31.62$, $72.66$, $130.88$, $207.15$, and $302.98$~meV. A constrained quadratic fit over the low-amplitude interval $A\leq0.20$~\AA{} yields $c_{\mathrm{DFT}}=3.2568$~eV/\AA$^2$,differing from the harmonic prediction by approximately $1.1\%$ and corresponding to $f_{\mathrm{DFT}}\simeq1.9325$~THz. At the largest distortion, $A=0.30$~\AA{}, the DFT energy is $302.98$~meV compared with the harmonic prediction $289.96$~meV, i.e.\ an upward deviation of about $4.5\%$. The amplitudes used for the main frozen-phonon analysis, in particular $A=0.20$~\AA{}, therefore remain close to the harmonic regime, while mild anharmonicity becomes visible only at the largest MnTe distortions.

The comparison is summarized in Table~\ref{tab:SM_harmonicity}. The
frequency inferred from the DFT curvature is obtained from
$f_{\mathrm{DFT}}=f_{\mathbf K}
\sqrt{c_{\mathrm{DFT}}/c_{\mathrm{harm}}}$.

\begin{table}[!htbp]
\centering
\caption{Consistency check between the phonon frequency and the curvature of
the frozen-phonon DFT total energy. For CrSb the quadratic fit uses the full
range $0\leq A\leq0.30$~\AA{}, whereas for MnTe the quoted DFT curvature is
fitted in the low-amplitude regime $A\leq0.20$~\AA{}.}
\label{tab:SM_harmonicity}
\begin{ruledtabular}
\begin{tabular}{lccccc}
System &
$f_{\mathbf K}$ (THz) &
$M_{\mathrm{eff}}$ (u) &
$c_{\mathrm{harm}}$ (eV/\AA$^2$) &
$c_{\mathrm{DFT}}$ (eV/\AA$^2$) &
$f_{\mathrm{DFT}}$ (THz) \\
CrSb & 3.1397 & 97.6747 & 8.8643 & 8.8966 & 3.1454 \\
MnTe & 1.9221 & 94.7208 & 3.2218 & 3.2568 & 1.9325 \\
\end{tabular}
\end{ruledtabular}
\end{table}

\section{Electronic structure of altermagnetic MnTe}

\begin{figure}[!htbp]
\centering
\includegraphics[width=0.84\linewidth]{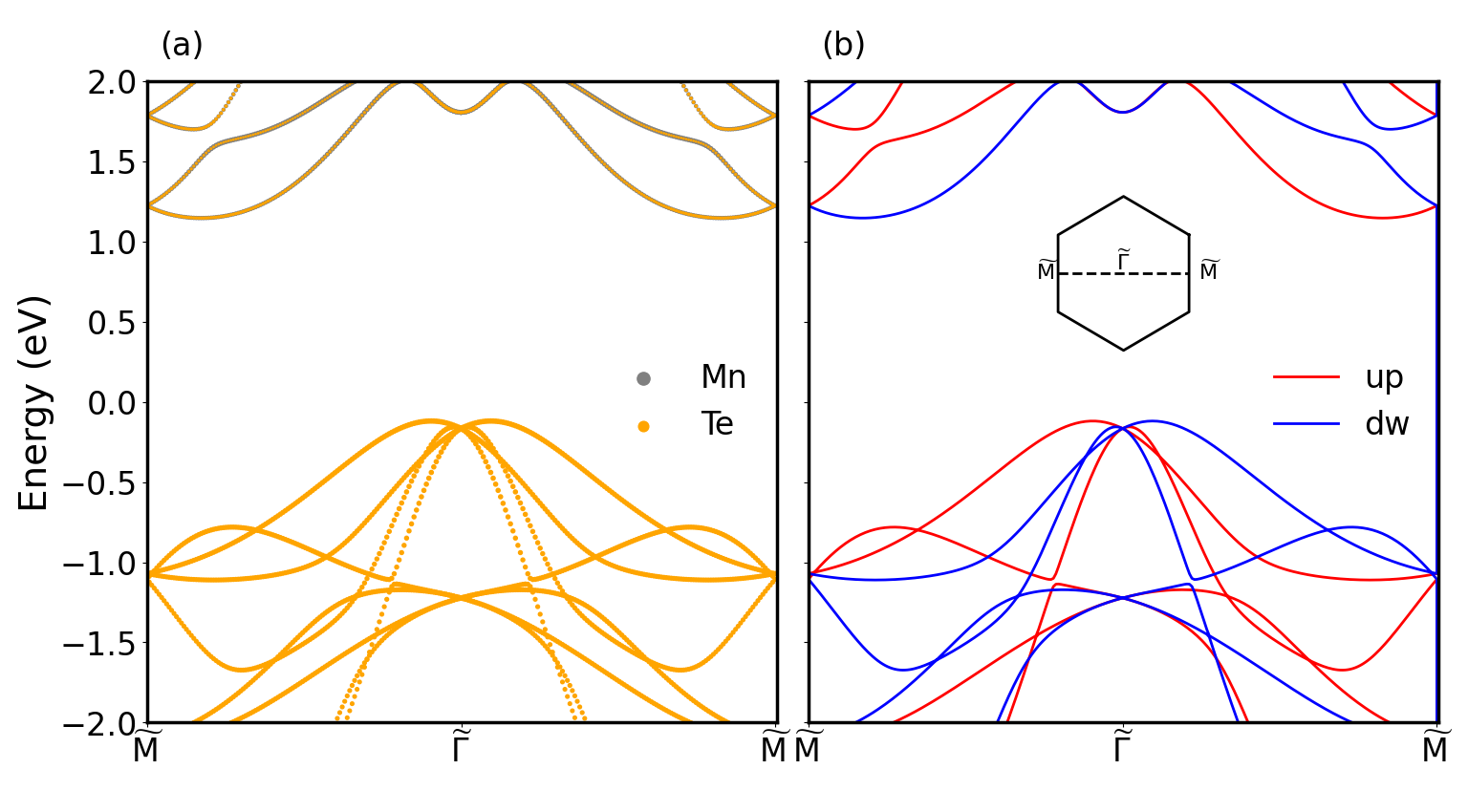}
\caption{Electronic structure of altermagnetic MnTe. (a) Orbital-resolved electronic band structure highlighting the Mn-$d$ contribution. (b) Corresponding spin-resolved electronic band structure showing the momentum-dependent spin splitting characteristic of the altermagnetic phase. The inset indicates the reciprocal-space path used for the electronic bands.}
\label{fig:mn_bands}
\end{figure}

Figure~\ref{fig:mn_bands} shows the electronic structure of MnTe in its compensated altermagnetic phase. As in CrSb, the magnetic order is associated with antiparallel moments on the two transition-metal sublattices, resulting in vanishing net magnetization while allowing momentum-dependent spin splitting of the electronic bands.

The orbital-resolved electronic structure shows that the states in the energy region relevant to the altermagnetic splitting contain a substantial Mn-$d$ contribution. The corresponding spin-resolved bands display the characteristic momentum-dependent spin splitting of the altermagnetic phase. The overall behavior closely parallels that obtained for CrSb in the main text and establishes MnTe as a second representative system in which spin-split altermagnetic electronic states coexist with valley-dependent lattice dynamics.

\section{Local chiral phonons in MnTe}

\begin{figure}[!htbp]
\centering
\includegraphics[width=0.75\linewidth]{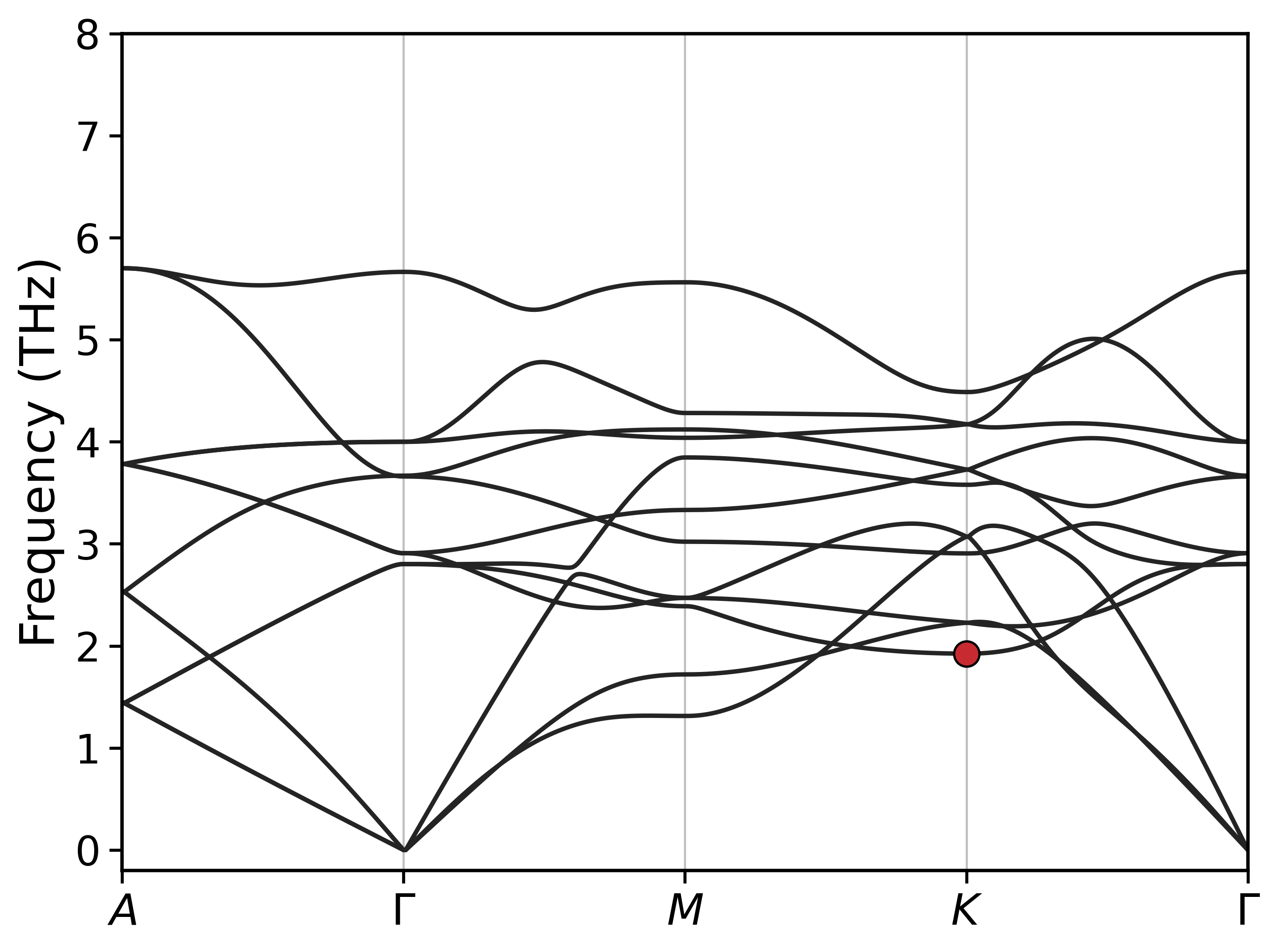}%
\caption{Phonon dispersion of pristine MnTe along
A-$\Gamma$-M-K-$\Gamma$. The highlighted point identifies the
lowest-frequency nondegenerate phonon mode at K
($f_{\mathrm K}\simeq1.9221$~THz), whose eigenvector is used for the
frozen-phonon distortions discussed in the main text.}
\label{fig:mn_phonons}
\end{figure}

The calculated phonon dispersion of MnTe is shown in Fig.~\ref{fig:mn_phonons}.
The selected lowest-frequency mode at K is nondegenerate
($f_{\mathrm K}=1.9221$~THz). Its eigenvector has out-of-plane
components on Mn and locally circular in-plane motion on the two Te sites.

Inspection of the corresponding eigenvectors shows that the local chiral lattice dynamics are predominantly associated with the Te sublattice. The two symmetry-related Te atoms undergo circular motion with opposite handedness and therefore carry finite but opposite local circular polarizations. Their contributions compensate in the pristine structure, resulting in a vanishing total phonon circular polarization despite the presence of locally chiral atomic motion.

This behavior mirrors that found for the Sb sublattice of CrSb and demonstrates that the separation between the atomic origin of the altermagnetic electronic structure and that of the local chiral lattice dynamics occurs in both compounds.

\section{Momentum-space texture of the local phonon chirality in MnTe}

\begin{figure}[!htbp]
\centering
\includegraphics[width=0.84\linewidth]{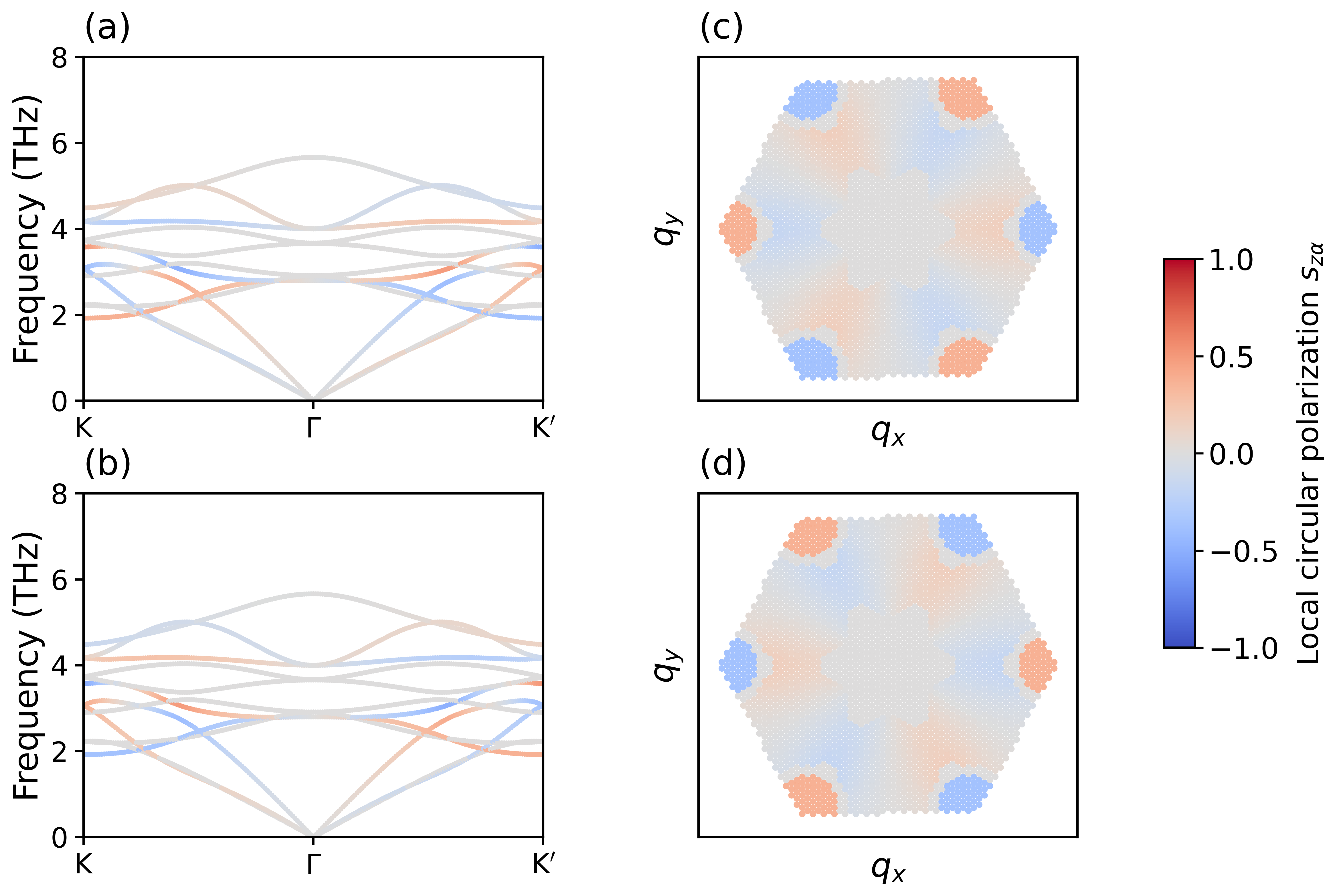}
\caption{(a,b) Local phonon circular polarization projected onto the two symmetry-related Te atoms of MnTe along the K--$\Gamma$--K$'$ path. (c,d) Momentum-resolved circular polarization projected onto the two Te atoms in the $q_z=0$ plane for the selected phonon branch identified in Fig.~\ref{fig:mn_phonons}. The alternating six-lobe pattern reveals the $f$-wave-like texture of the local phonon chirality.}
\label{fig:mn_chirality}
\end{figure}

Figure~\ref{fig:mn_chirality}(a,b) shows the atom-resolved circular polarization projected onto the two symmetry-related Te atoms along the K-$\Gamma$-K$'$ path. The two contributions have opposite signs and compensate when summed over the primitive cell.

Despite the vanishing total circular polarization, each Te sublattice exhibits a pronounced momentum dependence. The corresponding two-dimensional maps in the $q_z=0$ plane, shown in Fig.~\ref{fig:mn_chirality}(c,d), display an alternating six-lobe pattern analogous to that obtained for CrSb. The local phonon circular polarization therefore develops an $f$-wave-like momentum-space texture.

The opposite circular polarizations of the two Te sites demonstrate that the absence of a net phonon angular momentum in pristine MnTe does not imply the absence of local chiral lattice dynamics. Instead, the chirality is hidden by symmetry-enforced compensation between the two atomic contributions.

\section{Frozen-phonon response of MnTe}

The lowest-frequency nondegenerate phonon at K was used for the MnTe
frozen-phonon calculations. The structural snapshots and the electronic
band unfolding follow Eqs.~\eqref{eq:SM_Meff}--\eqref{eq:SM_frozen}.
The resulting representative band separations are shown in Fig.~5 of the
main text; the same panel is not reproduced here. At $A=0.20$~\AA,
the corresponding maximum Mn and Te displacements are $0.0951$~\AA{}
and $0.0740$~\AA, respectively. The leading response is approximately
linear in the structural amplitude, with local separations rather than
global band gaps being extracted.

\section{Symmetry breaking and valley phonon chirality in Mn$_2$TeSe}

We finally consider the effect of 50\% isoelectronic Te$\rightarrow$Se substitution in MnTe. The resulting ordered Mn$_2$TeSe structure removes the equivalence between the two chalcogen sites and hence the symmetry responsible for the exact compensation of their local circular polarizations.

The substituted system retains the characteristic momentum-dependent spin splitting of the altermagnetic electronic structure. At the same time, the reduced symmetry allows the phonon spectrum to develop a finite total circular polarization at the K and K$'$ valleys. In contrast to pristine MnTe, for which the atom-resolved contributions compensate within each valley, the symmetry-broken structure can exhibit finite and opposite valley polarizations,
\begin{equation}
s_z^{(\mathrm K)}=-s_z^{(\mathrm K')},
\end{equation}
and consequently a finite valley contrast $\Delta s_z$.

The 50\% substituted structure thus provides a simple ordered model demonstrating how the compensated local circular motion of pristine MnTe can be converted into an uncompensated valley-resolved phonon angular momentum.
\begin{figure}[!htbp]
\centering
\includegraphics[width=0.95\linewidth]{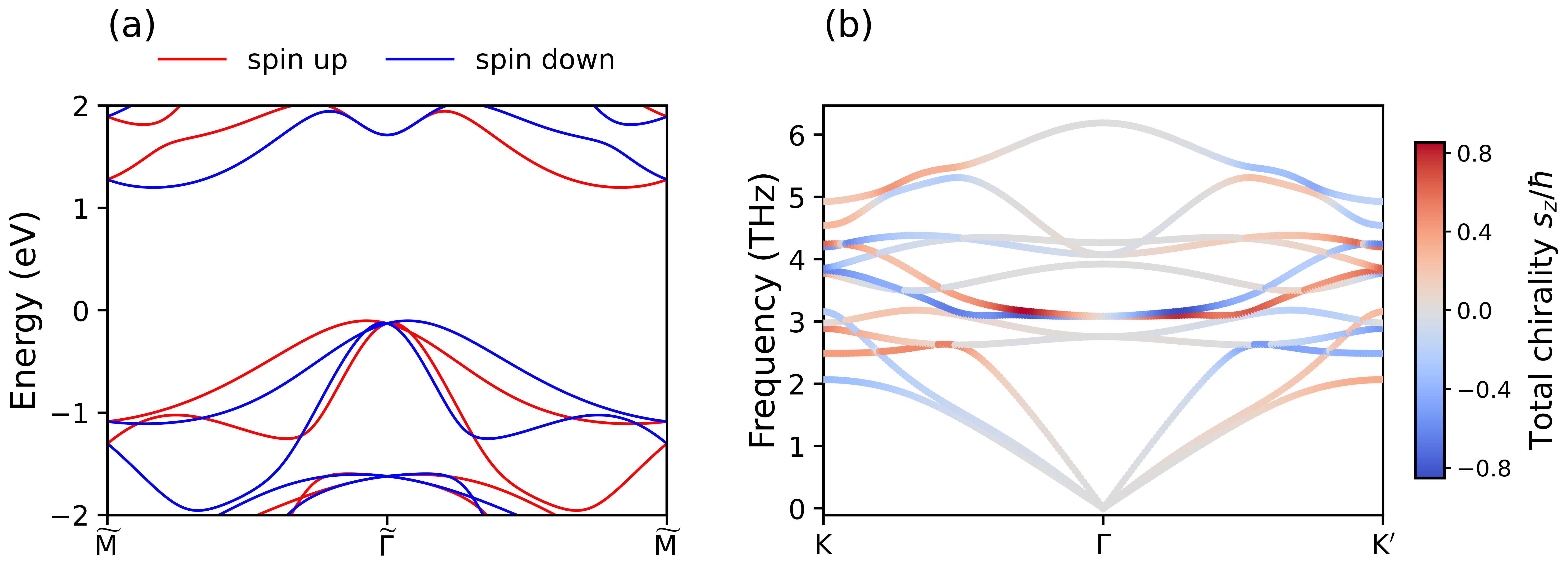}
\caption{(a) Spin-resolved electronic band structure of Mn$_2$TeSe, showing the persistence of momentum-dependent spin splitting after 50\% Te$\rightarrow$Se substitution. (b) Phonon dispersion along K-$\Gamma$-K$'$ colored by the total phonon circular polarization $s_z$, illustrating the emergence of finite valley phonon chirality after symmetry breaking.}
\label{fig:mn_te_se}
\end{figure}

\bibliography{bibliography_SM}